\documentclass[preprint,showpacs,showkeys,aps,prb]{revtex4}
\usepackage[english]{babel}

\usepackage[dvips]{graphicx}
\usepackage{amssymb}

\begin{document}


\title{50 Years of Computer Simulation --- a Personal View}

\author{Wm. G. Hoover\\
Ruby Valley Research Institute \\ Highway Contract 60,
Boxes 598 and 601, Ruby Valley 89833, NV USA \\
     }
\date{\today}




\vskip 0.5cm

\begin{abstract}

In the half century since the 1950s computer simulation has transformed our
understanding of physics. The rare, expensive, slow, and bulky mainframes of
World War II have given way to today's millions of cheap, fast, desksized
workstations and personal computers.  As a result of these changes, the
theoretical formal view of physics has gradually shifted, so as to focus on
the pragmatic and useful.  General but vague approaches are being superceded
by specific results for definite models.  During this evolving change of
emphasis I learned, developed, and described my simulation skills at
Michigan, at Duke, at Livermore, and in Nevada, while forming increasingly
wide-ranging contacts around the world.  Computation is now pervasive in
all the scientific fields.  My own focus has been on the physics of
{\em particle} simulations, mainly away from equilibrium.  I outline my
particle work here.  It has led me to a model-based understanding of both
equilibrium and nonequilibrium physics. There are still some gaps.  There
is still much to do.

\end{abstract}

\maketitle

\section{Introduction}

Computer-induced changes in emphasis have transformed what it means to
``understand'' physics.  This transformation is nowhere more striking
than in the many-body model-based subjects of statistical mechanics
and kinetic theory.  The {\em old} way was solving many-body problems ``in
principle'' (but not ``in fact''), by formal expansions around the ideal
gas or the harmonic crystal.  The {\em new} way has replaced the
expansions with direct numerical simulations of model systems. At
equilibrium, ensemble-based ``Monte Carlo'' simulations are appropriate.
``Molecular dynamics'' simulations are more generally useful because
they  apply to both equilibrium and {\em nonequilibrium} situations.
The nonequilibrium problems typically involve flow and gradients and
are quite unlike their static homogeneous equilibrium relatives.
Writing down the governing equations for a many-body model (a sum over
states, or the dynamical equations of motion, for instance) is now
considered just a beginning challenge, not an end in itself.

Masaharu Isobe thought my reminiscences would interest younger readers
of this Journal, ``Ensemble''.  Besides the name's possible link to
Gibbs' ensembles, the name ``Ensemble'' also suggests
cooperation, and ``working together''.  These are nice concepts, which
underlie the steady progress of science, particularly in today's
electronic world, with the immediacy of the internet and email making
efficient timely international collaborations possible.

I agreed with Masaharu's idea, and summarize some milestones of my work
here.  Though times and tastes change, the experiences of learning, 
creating, and teaching trace an enduring continuity worth considering
and summarizing.  Here is a bit of my own simulation history, with the
hope it will prove useful to yours.  I start out with my college and
university days in Ohio and Michigan, and end up with retirement in Nevada.
Many more details can be found on my website [ http://williamhoover.info ].

\section{Ohio, 1953-1958, and Michigan, 1958-1961}

My small and isolated undergraduate college, Oberlin College in Ohio,
featured and encouraged the sceptical attitude so useful in physics.
Otherwise the scientific content of my Oberlin liberal-arts coursework
was irrelevant to my research career.  The classroom instruction
in mathematics and physics was formal, and mired in the past. This
disappointing style of instruction very nearly convinced me to switch from
science to economics, my Father's field of study.

Recuperation from an automobile accident kept me out of college for
a semester.  I attended Harvard's Summer School to make up the
Physical Chemistry course I had missed.  The lecturer, Stuart Rice,
provided  research excitement and strengthened my commitment to science.
Stuart loved kinetic theory.  He reassured us students that any
shortcomings in the practical labwork portion of his course would have
no influence on our grades.

My graduate university, the University of Michigan in Ann Arbor, was a
great improvement over Oberlin.  The University of Michigan provided
useful coursework,
state-of-the-art computational tools, as well as the stimulation and
inspiration necessary to research.  Shortly after my arrival there I
came upon Alder and Wainwright's ``Molecules in Motion'' article in
Scientific American\cite{b1}.  The so-different pictures  of hard-sphere fluid
and solid trajectories in that article are still worth a look today.
See {\bf Figure 1} for a pair of example pictures taken from Farid
Abraham's 1980  work on the melting transition\cite{b2}.  In the 1950s,
simulating the motion of 108 particles was a challenge.  The
Alder-Wainwright pictures triggered my interests in microstructure and
molecular dynamics.

\begin{figure}
\includegraphics[height=12cm,width=14cm,angle=-0]{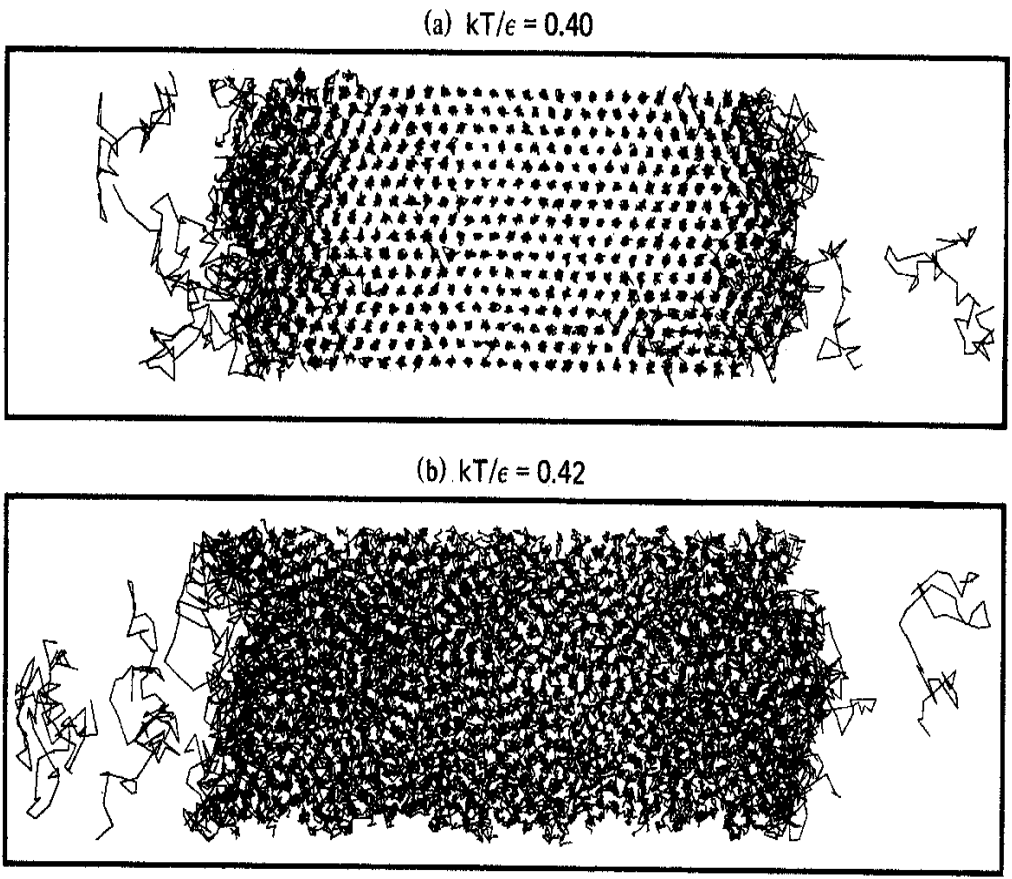}
\caption{
Atomistic trajectories just below (top) and just above (bottom) the
melting temperature, taken from Reference 2.
}
\end{figure}

At Michigan, Andrew Gabriel De Rocco, a young Chemical Physics Professor
and my PhD thesis advisor, was excited by statistical mechanics.  He had a
rare knack for relating formalism to the real world.  What constitutes the
``real world'' is of course a matter of opinion.  I well remember Andy's
wife Sue, who overheard one of our upstairs conversations on pair
potentials, shouting up the stairs: ``Andrew, {\em all} your potentials are
repulsive!''  Tastes differ!

As a result of Andy's enthusiasm and support
I became an expert in the Mayers' ``virial'' expansion, the expression of
fluid pressure as a power series in the density.  $B_n$, the $n$th
coefficient in the series is a sum of relatively-complicated $n$-body
integrals.  The integrands of these integrals are products of as many
as $n(n-1)/2$ functions linking the $n$ particles together.  For hard
parallel squares and cubes these integrals can be done
analytically\cite{b3}, though beyond $B_6$, the topological bookkeeping
requires a fast computer.  By 1960 I had the patience and the training to
program the FORTRAN calculation of a few million of these integrals,
using the ``MAD'' computer [Michigan Algorithmic Decoder]\cite{b4}.
Programming was then a bit tedious.  It involved writing instructions in
the form of punched cards, one card for each program line. But it had
to be done.  For parallel cubes $B_7$ required computing
$468 \times 7 ! = 2,358,720$ separate integrals!  MAD made occasional
irreproducible machine errors.  These errors were useful reminders of the
need for vigilance in numerical work.  The violation of obvious
requirements (such as conservation of momentum and energy) is the
usual result of logical or typographical errors in programming.

Once the program was successfully punched out, and carefully checked,
I found {\em negative} virial
coefficients, both $B_6$ and $B_7$, for hard {\em repulsive} parallel
cubes.  Negative tensile contributions for positively repulsive particles
was a big surprise!  I was thoroughly hooked by the excitement of research.
Of course the pace was slower then.  I exchanged several letters with
``H. N. V.'' Temperley about the details of the virial series. After a few
letters the ``H. N. V.'' changed to an informal ``Neville''. A typical
US $\leftrightarrow$ UK roundtrip correspondence took two weeks by airmail.
Bob Zwanzig had published the paper\cite{b3} which introduced me to hard
cubes but which also wrongly contradicted my negative-coefficient results.
It was a real thrill, and a lesson, when, in a phone conversation with Andy
and me, he readily admitted his mistake.

\section{Duke University, Durham, North Carolina, 1961-1962}

In the 1960s in America a postdoctoral appointment was required before
seeking out a ``real job''.  Andy sent me on to Duke University, where
one of John Kirkwood's students, Jacques Poirier, lived
an isolated existence as a Professor of Chemistry with theoretical
inclinations.  The smell of menthol from the Salem cigarette
plant traveled miles in the evening to our house on a dirt road next to
a turnip field. The various organic smells during the day at the Chemistry
Department helped make this a
tranquil detour from mainstream physics.  Because Jacques' ideas for our joint
research turned out to be invalid I was able to pursue further work
on the virial series while at Duke.  In those days, when computer
simulation was still rare, approximate integral equations for the
distribution of particle pairs $g(r)$ were all the rage.  Because the
equations were nonlinear in $g(r)$, complete solutions required elaborate
computation.  But substituting a density expansion of $g(r)$, and
equating coeffients of powers of the density made it possible to
compute the approximate virial coefficients and compare them to the
Mayers' exact expressions\cite{b5}.  

During the Duke year, I got in contact with George Stell, another
{\em afficianado} of the Mayers' series, a skilled jazz musician, and still
a great friend.  I visited George in his Greenwich Village apartment.
Two memories of that visit stand out: George had his own sauna there,
and the relative calm in his apartment was broken by a sewergas
explosion just outside, strong enough to levitate a heavy manhole cover.
Once back at Duke I managed to land job interviews at the Livermore
and Los Alamos Laboratories, the two rival computing giants overseen
by the University of California.  Alder and Wood, at the two
laboratories where I would soon seek a job, were rivals too,
disagreeing over their relative priority and contributions to the
understanding of the hard-disk and hard-sphere melting
transitions\cite{b6,b7}.

\section{Livermore and Davis, California, 1962-2004}

I interviewed at both Los Alamos and Livermore in 1962, flying out from
Michigan to talk to a dozen or so scientists at each laboratory and
giving a seminar on my virial-series work.  Los Alamos, in the
mountains of New Mexico, was physically the more interesting of the two
bomb laboratories.  Livermore was located in a once rural grape and cattle
ranching valley suffering now from pollution and urban sprawl.  By the
time I arrived Alder and Wainwright were studying disk and sphere systems
of about 1000 particles, using both rigid and periodic boundary conditions.
One of their research goals was understanding whether or not disks and
spheres could freeze and melt at high density.  My hard-particle virial
series expertise fitted in well with that work\cite{b8}.

At Los Alamos the relative humidity was low and the scientists wore
Hawaiian shirts and desert boots rather than the suits of the Midwest.
The salaries there were nearly 20 percent lower (partly a function of
the longer five-week vacations due to the laboratory's remoteness).
This difference in salaries decided me on Livermore, where Berni Alder
provided me a home in the Physics Department.  It was an exciting time
and place. Particle, continuum, plasma, astrophysical and
nuclear physicists all joined together, with weekly scientific
meetings under Edward Teller's watchful eyes.  For one of the talks
Francis Ree and I got Teller's permission to pursue and present a
computationally demanding single-occupancy explanation of the
hard-disk and hard-sphere phase transitions\cite{b9}.  Francis Ree,
trained by Henry Eyring, was, like me, dedicated to precise
and careful statistical analyses.  We solved Berni Alder and Bill Wood's
hard-sphere and hard-disk problems using both the Mayers' virial series
(for the fluid phases) and Metropolis-Rosenbluth-Teller Monte Carlo
simulations (for Kirkwood's single-occupancy solid phases).   See
{\bf Figure 2}.
Our ``Monday Morning Meeting'' presentation went well.

\begin{figure}
\includegraphics[height=12cm,width=12cm,angle=-0]{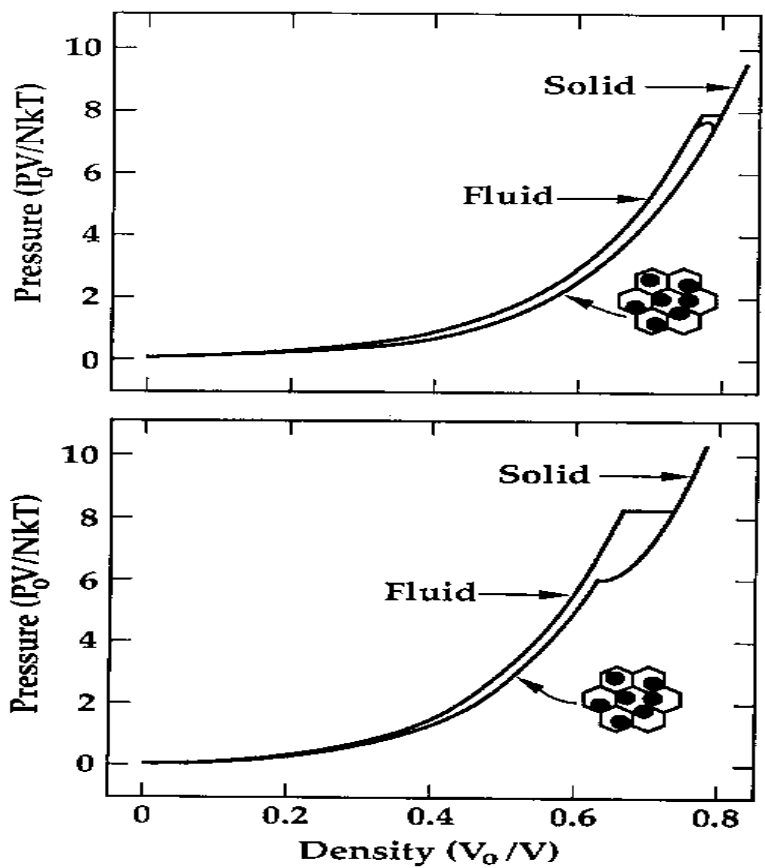}
\caption{
Hard-disk (above) and hard-sphere (below) equations of state. From
Reference 9.  The close-packed volume (area in two dimensions) is $V_0$.
The lower ``single-occupancy'' curves correspond to the pressure of
particles confined to space-filling cells, as shown in the insets.
}
\end{figure}

This Monday-Morning-Meeting background in a variety of disciplines was
extremely useful to me later on, both scientifically and socially.  Teller
started the Department of Applied Science at Livermore in 1963, with the
idea that a wide background (nuclear, quantum, classical, mathematical,
chemical, electromagnetic) was essential to training students.  I taught
in the Department for about thirty years, and enjoyed it thoroughly.  My
wife Carol was a student in one of the statistical mechanics courses I
taught at DAS.

The bookkeeping associated with hard-particle collisions made molecular
dynamics seem less interesting to me (or at least harder) than Monte Carlo
simulation.  But in the early 1960s Rahman, Verlet, and Vineyard\cite{b10}
all followed Fermi's 1950s work, in showing how to carry out simulations
with continuous force laws. My group leader at the time, Russ Duff,
supported my interest in learning molecular dynamics.  As a shockwave
experimentalist he liked the idea of simulating  shockwave-induced
melting with molecular dynamics.  I took on the project, storing thousands
of long particle trajectories on magnetic tapes.  The physical stretching
of some of these tapes made them unreadable often enough that the
storage procedure was useless for long runs.  The shockwave simulations
of 1967 had to be put aside temporarily, and were not resurrected until
1980\cite{b11}, when magnetic tapes were obsolete, and again in
1997\cite{b12}, when the anisotropicity of temperature had caught my
interest.  {\bf Figure 3}
illustrates that interesting result from the shockwave work: in strong
shockwaves temperature becomes a tensor.  The longitudinal temperature
(the velocity fluctuation in the direction of the motion) greatly exceeds
the transverse temperature in strong shocks.

\begin{figure}
\includegraphics[height=12cm,width=16cm,angle=-0]{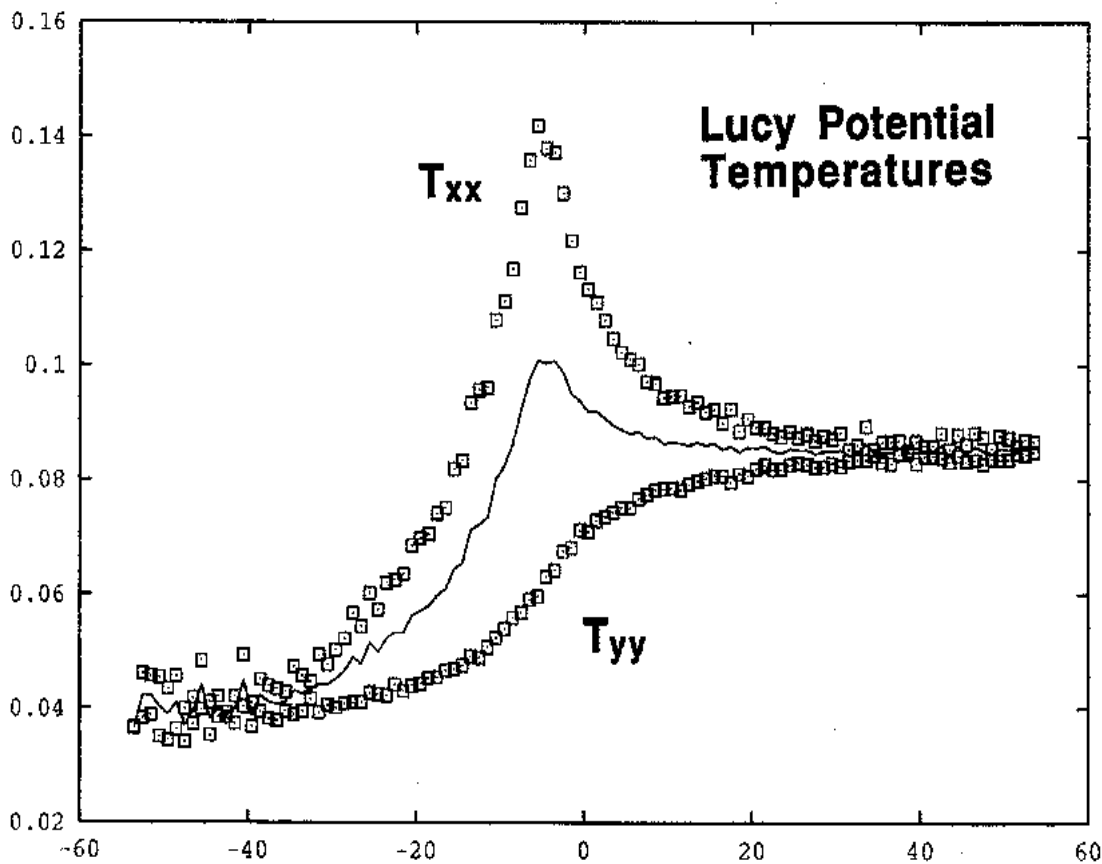}
\caption{
Longitudinal (top), mean, and transverse (bottom) kinetic temperatures
in a Lucy fluid shockwave, taken from Reference 12.
}
\end{figure}

A later group leader of mine, Mark Wilkins, strongly influenced my
scientific outlook.  Mark was a self-taught expert in the numerical
solution of the partial differential equations of continuum mechanics,
specializing in plastic flow and fracture, essential for weapons
simulations.  Mark stressed that physics is a study of ``models'', 
closed sets of differential equations which bear some resemblance to
our experiences in ``real life''.  Quantum mechanics is such a model,
and evidently quite imperfect, in that it never predicts the unique
outcomes observed in the real world.
 
At the Livermore Laboratory of the 1960s there were plenty of interesting
collaborators with whom I could work.  Al Holt was a formalist, a
tensor specialist who had been trained by chaos guru Joe Ford at
Georgia Tech.  Al's interest in elastic constants combined nicely with
my own statistical and lattice dynamical skills.  Dave Squire worked
for the Army Research Office in Durham, North Carolina.  He was a
practical chemist, schooled by Zevi Salsburg, a colleague of Berni's.
Dave was able to visit the
Livermore Laboratory for several months in the late 1960s, along with
his wife and eight children.  Al and Dave and I formulated the
elastic response of crystals to strain under both isothermal\cite{b13} and
adiabatic\cite{b14} conditions.  These results, which can be evaluated by
either Monte Carlo or molecular dynamics simulations, have been rediscovered
frequently.

I visited the nearby IBM Research Laboratory at Almaden, which had
managed to hire John Barker and Doug Henderson.  I was invited there
for a seminar and prepared a set of slides.  Barker ran the slide
projector while Henderson asked an occasional question.  That was the
whole audience at Almaden.  I was fortunate, at Livermore, that dozens
of scientists were actively doing research in statistical mechanics.
In the late 1960s Barker and Henderson, along with Mansoori, Canfield,
Weeks, Chandler, Andersen, Rasaiah, and Stell, developed a very
successful perturbation theory for the Helmholtz free energy\cite{b15}.
Only the hard-sphere pair distribution function was needed, and
a useful form for that was available from the integral-equation work. 
The theory was actually useful for realworld equilibrium thermodynamic
calculations.  This equilibrium breakthrough convinced me it was high
time to switch to the study of {\em nonequilibrium} problems, where the
only theory available was Green and Kubo's linear-response theory of
transport\cite{b16}.

The energy crunch in the Carter administration led to mass firings at
the Livermore Laboratory and to a change of emphasis and structure: 
suddenly there were lots of group leaders, lots of progress reports
and proposals, and detailed budgets.  When I needed dozens of hours of
CRAY computer time to study the effect of Coriolis' forces on the heat
flux, Roger Minich, a favorite of one of the bomb divisions, generously
gave me the time from his weapons-physics accounts\cite{b17}.

The cutback in basic research at Livermore made it necessary to look
outside the laboratory for collaborators.  I had a lot of fun working
with Brad Holian at Los Alamos.  We had a common interest in
statistical mechanics and simulation.  Brad tried to get a job at
Livermore; I tried to get a job at Los Alamos, so that we could work
together, but both these initiatives were unsuccessful.

In my attempts to satisfy the mounting laboratory pressure for
``relevance'' I carried out some dynamic fracture simulations with Bill
Ashurst\cite{b18} and, in his PhD thesis work, Bill Moran.  {\bf Figure 4}
shows a typical fracture specimen.  At a meeting with the Laboratory
Director, Mike May, Mike asked me whether or not these fracture
simulations were really relevant.  I had to admit that atomistic models
are actually quite limited in scope, and are often misleading (even today).

\begin{figure}
\includegraphics[height=8cm,width=16cm,angle=-0]{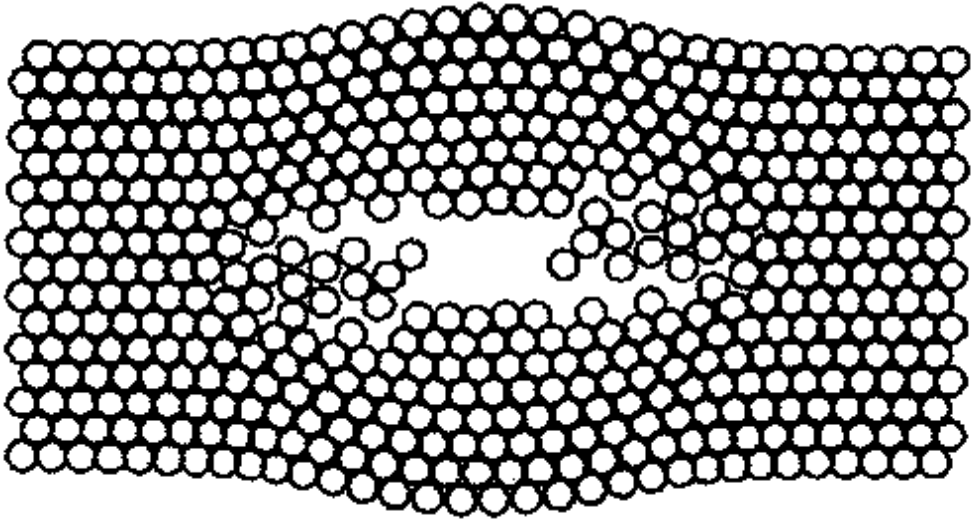}
\caption{
Fracture specimen simulation, taken from Reference 18.
}
\end{figure}

Bill Ashurst, my first PhD student at the Department of Applied Science,
was keen to simulate nonequilibrium flows.  We developed methods for
simulating shear flows with isothermal boundaries as well as periodic
homogeneous algorithms\cite{b19}.  Bill was able to make movies (as was
also Brad Holian at Los Alamos).  These movies were a fixture of small
topical physics meetings in the early days of nonequilibrium molecular
dynamics.

An early experience with the National Science Foundation was educational
and helped sharpen my scepticism for government's ability to select
good problems to solve.  In the early 1980s I
proposed a Fourier analysis of simple nonequilibrium distribution functions.
The proposal was evaluated ``excellent'' by five reviewers and ``very
good'' by the sixth.  Despite this, the proposal was declined.  This early
failure reinforced my natural inclination to forgo begging for research
funds.

\section{Travel, with Sabbaticals in Australia, Austria, and Japan}

After a first visit to France to confront Verlet, Levesque, and
K\"urkijarvi's somewhat erroneous Green-Kubo transport
results\cite{b20} with the nonequilibrium analogs I had calculated 
with Bill Ashurst, Orsay became a regular stop for me.  Carl Moser
supervised a long series of workshops, seminars, and
meetings, at CECAM (The European Center for Atomic and Molecular
Simulation) which were seminal and stimulating.  The locale, about an
hour outside Paris in the countryside, was conducive to good work.
There was a hillside covered with wild blackberries on the way to lunch
and the Parisian bistros and restaurants attracted us in the evenings
after work.

The Department of Applied Science and the Livermore Laboratory
provided my financial support throughout each year.  Though my salary
was set by the Department, based on academic criteria, the Laboratory
always tried to exert pressure toward ``practical applications'' to
justify its paying the Lion's Share (five eighths) of my salary.  As
compensation for this pressure the ``Professional Research and Teaching
Leaves'' available at the Laboratory made it possible to get away for
sabbatical research.  Such leaves, augmented by support from the
Fulbright Foundation, Universit\"at Wien, and the Japan Society for
the Promotion of Science, took me and my family to Australia in 1977,
to Austria in 1984, and to Japan in 1989.

In 1980, between my Australian and Austrian sabbaticals, I noticed
that a Hamiltonian could be constructed which reproduced exactly the
geometry and energy balance for a many-body system undergoing periodic
shear.  This was the ``Doll's Tensor" Hamiltonian, which I described in
June 1980 at Sitges\cite{b21} (Spain) after the StatPhys organizers for the 
August 1980 Edmonton meeting turned down my proposal to speak about it in
Canada.  ``Doll's Tensor'' was simply the tensor $qp$ array, constructed
of the Cartesian particle coordinates $\{ q\}$ and momenta $\{ p \}$.
Adding the term
$ \dot\epsilon\sum yp_x$ to the usual many-body Hamiltonian induces
a macroscopic motion with $\langle \dot x \rangle = \dot \epsilon y$,
simple shear, or ``Plane Couette flow''.  In June 1982, at Howard Hanley's
seminal meeting, ``Nonlinear Fluid Behavior'', which I helped organize,
I was finally able to present a talk on nonequilibrium molecular dynamics
to a large mostly-American audience of interested colleagues\cite{b22}.

By 1984 my desire to present the Doll's-tensor work as a talk at an
international conference, the 1983 Edinburgh StatPhys meeting, had 
once again been frustrated.  The consolation prize, a humorous poster
detailing the history of the shear flow work, was not published until
nine years later in the proceedings of a Sardinia meeting\cite{b23}.
In reaction to all the StatPhys frustration, I organized, with
Giovanni Ciccotti, a highly-successful
Enrico Fermi Summer School meeting at Lake Como, where the new
nonequilibrium algorithms were thoroughly discussed\cite{b24}.
See particularly Denis Evans' lecture ``Nonequilibrium Molecular
Dynamics'', pages 221-240 of the School's Proceedings.

Only very recently\cite{b25}, with my wife Carol and Janka Petravic, I
have quantified the errors (nonlinear in the strainrate $\dot\epsilon $)
incurred by using the Doll's and the closely related S'llod algorithms.
The kewpie doll has a highly-interesting history (the Centennial of
the Doll is 2009!) in addition to its usefulness as a
mascot for statistical physics.

With the shearflow problem solved, Denis Evans and Mike Gillan, working
completely independently, found an external field that correctly
generated heat flow in 1982 -- see again Evans' lecture\cite{b24} for the details.
Their solution of this problem was particularly interesting because it
provided a concrete example in which Gauss' Principle (equivalent to
Least Action for equilibrium systems) gives incorrect motion equations
(inconsistent with Green-Kubo) away from equilibrium\cite{b26}.

The Australian Sabbatical experience had been interesting, though
slightly chaotic.  My proposed work at the
Australian National University's Computer Centre, in Canberra, with
Bob Watts, came to an abrupt end in the first week when Bob was
appointed to replace the Director of the center.  Watts' water
potential, which I had intended to investigate in Australia, turned
out to be unstable, making it possible to concentrate on what was for me
a more interesting project, the determination of liquid and solid free
volumes.  My son Nathan, having just finished high school, was in 
Australia with me and we worked together at the ANU Computer Centre.
That work involved a ``gedanken experiment'' in which a single
very light particle traced out a ``free volume'' while its heavier
neighbors stayed put\cite{b27}.  I had used this same idea earlier
to rationalize the use of cell models\cite{b28}.  {\bf Figure 5} illustrates
the difference between the fluid and solid phases from this perspective.
It was educational to learn from these results that the free volume in
the fluid phase is actually smaller than that in the coexisting solid!

\begin{figure}
\includegraphics[height=8cm,width=16cm,angle=-0]{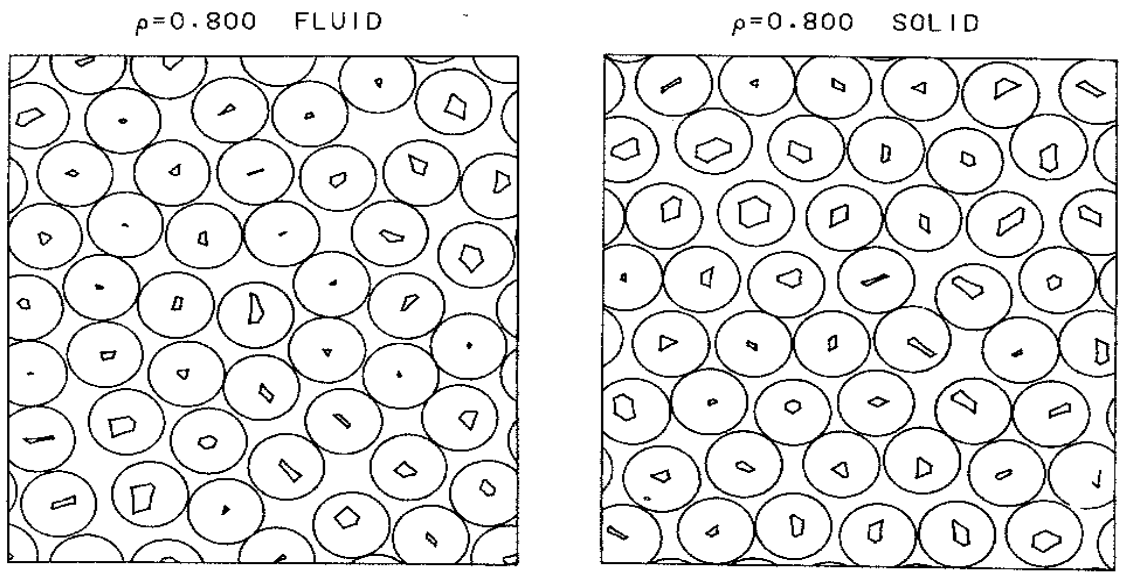}
\caption{
Fluid (left) and solid (right) hard-disk free volumes at a common density
four-fifths of close packing.  The ``free volumes'' shown are the regions
available to the center of each disk when the remaining disks are held
fixed.  Taken from Reference 27.
}
\end{figure}

My first Austrian experience also had an unexpected turn.  Rather than working
exclusively with Karl Kratky, as I had intended, I began to collaborate
with Harald Posch\cite{b28}, whose interests in statistical mechanics and 
nonequilibrium simulations were similar to mine.  Both Harald and I valued
reproducibility and precision very highly.  We often compared results from
our two independently-written computer programs.  Kratky was a formalist,
and I soon lost patience waiting the months it took him to reproduce
results I could generate numerically in a matter of days.  The Austrian
sabbatical gave me background for my first book, ``Molecular Dynamics'',
lectures given at Universit\"at Wien and written up once I was back at
Livermore, in 1986.  In those electronically primative times it was necessary
to edit the teX file for the manuscript in one building and to drive about
a half mile to another building to see a printout.  Because car travel was
limited until 18:00 during the week, most of the book work had to be done
at night.

Prior to a 1984 CECAM workshop at Orsay I had the very good
fortune to meet Shuichi Nos\'e on the Orly Airport train platform.  (I had
noticed his surname printed on his suitcase.)
This meeting eventually led to a very pleasant and creative year in Japan.
Nos\'e was isolated from the other workshop members, choosing to stay
in a Japanese-style hotel.  We arranged to meet at the Notre Dame
cathedral.  On a bench in front of the church we talked about his
novel thermostat ideas\cite{b30,b31} in detail.  After the workshop I visited
Philippe Choquard in Lausanne to work out the consequences for a
harmonic oscillator.  The result was my most cited paper\cite{b32}. Some of
the oscillator orbits were quite beautiful\cite{b33}.  {\bf Figure 6} shows
a regular periodic orbit for a thermostated oscillator.  The oscillator
exhibits chaotic orbits too.  ``Chaotic'' orbits
have the property of Lyapunov instability -- an infinitesimal perturbation
of such an orbit grows exponentially fast with time.  Orbits for
dissipative systems, in which work is converted to heat, are typically
``fractal'' with a fractional dimensionality less than that of the
space in which they are embedded.  See {\bf Figure 7} for an example.

\begin{figure}
\includegraphics[height=12cm,width=12cm,angle=-0]{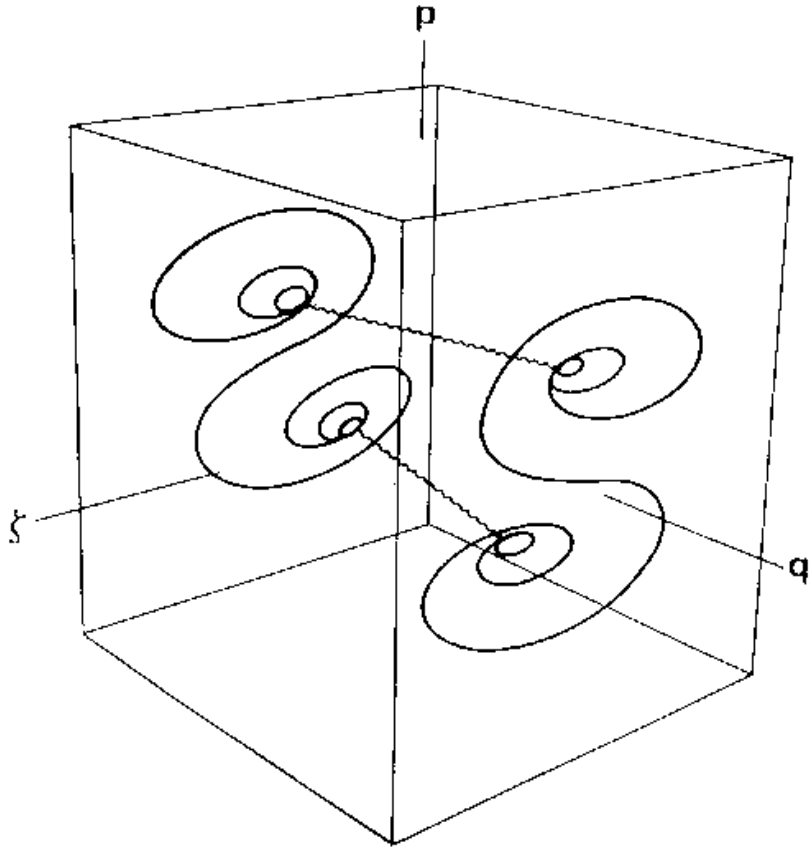}
\caption{
Periodic orbit for a thermostated harmonic oscillator with coordinate
$q$, momentum $p$, and friction coefficient $\zeta $, taken from
Reference 33.
}
\end{figure}

My interest in computational thermostats was immediate and has
continued to this day.  I asked Berni Alder what he thought about my
energy and temperature-control ideas.  He pooh-poohed the notion.  I
traveled to Los Alamos to ask Bill Wood for his ideas.  Though a bit
more diplomatic, his thoughts were the same as Berni's: thermostats
were not a very useful idea.  Fortunately, I had the freedom to spend
much of the next few years working out the details, linking thermostats
to statistical physics.  Bill Moran and I generated the fractal objects
which describe the collisions taking place for a very simple problem.
We studied a thermostated particle falling through a periodic ``Galton
Board'' of hard-disk scatterers\cite{b34}.  The collisions formed a
multifractal object, with the dimensionality of the object decreasing
with increasing field strength.  For a sample see {\bf Figure 7}.  

\begin{figure}
\includegraphics[height=12cm,width=12cm,angle=-0]{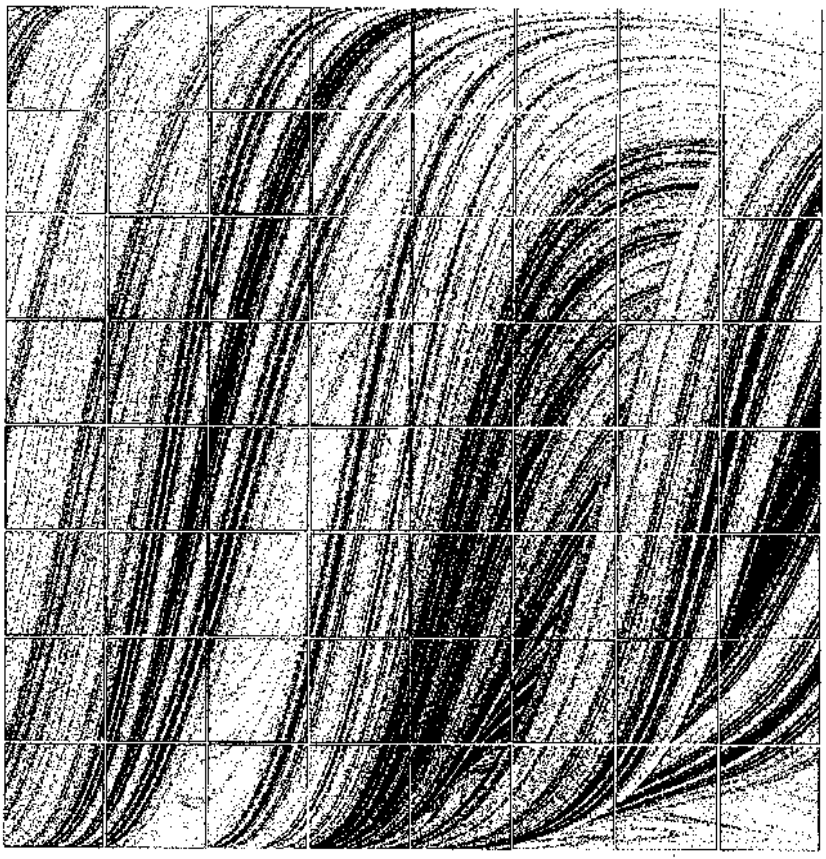}
\caption{
A multifractal phase-space plot of successive collisions in the Galton
Board problem with a constant vertical gravitational field.
The abscissa corresponds to the location of each collision on a hard
disk scatterer from bottom (at the left) to top (at the right) relative
to the downward field direction.  The ordinate corresponds to the tangential
velocity component, which varies from $-1$ (at the bottom) to $+1$ 
(at the top).  The motion of the moving particle is constrained to a
constant speed of unity, as explained in Reference 34.
}
\end{figure}

The finding that the dimensionality of phase-space distributions was
reduced below that of Gibbs' equilibrium distribution was a revealing
and rewarding insight for me.  The extreme rarity of the fractal
nonequilibrium states explained irreversibility.  Because the equations
of motion are time-reversible and the probability of choosing an initial
fractal state is of measure zero, the probability of {\em violating} the
Second Law of Thermodynamics, along a time-reversed trajectory, vanishes.
The fractal nature
of the phase-space distributions also showed that there could be no
nonequilibrium entropy.  This is because Gibbs' recipe for the entropy in terms
of the $N$-body distribution function $f$ and Boltzmann's constant $k$,
$$
S_{\rm Gibbs} \equiv-k\langle f\ln f \rangle \ ,
$$
{\em diverges} when $f$ is singular.

At the New York American Physical Society
meeting emphasizing the hot topic of ``High-Temperature
Superconductors'' I took long walks in Central Park, mentally estimating
the phase-space dimensionality loss in strong shockwaves and ruminating
over the paradox that time-reversible equations of motion lead to
irreversible behavior.  Much later these topics gave rise to my third
book, ``Time Reversibility, Computer Simulation, and Chaos''.  Oddly
enough, much of today's research still deals with equilibrium problems,
though to me nonequilibrium ones are more numerous, more significant,
and more interesting.

In the years following my 1977-1978 sabbatical Down Under,
nonequilibrium molecular dynamics had been developing rapidly.  In this
same period, my
marriage was deteriorating.  The years from 1980 through my divorce,
in 1986, were particularly difficult, but still relatively
productive.  In 1988 a very lucky chance meeting
with a visitor to the Livermore laboratory, a young researcher from
Louisiana State University, a Doctor Gupta, brought me
in contact with his host, a former student of mine, Carol Griswold
Tull.   Carol was working with the Livermore supercomputers and her
own marriage had ended in 1984.   We were fortunate to share very similar
interests in a ``Good Life'' mixture of science, nature, music, and
nourishment.  At last I had found a faithful woman with whom to share
my life.  Our marriage was arranged for 1989, so that we would arrive
for our sabbatical in Japan as an officially married couple.  My son
Nathan performed the ceremony in Carol's Livermore home.

The Japanese experience, at Keio University in Yokohama, was a surprise.
After finding that there seemed to be no plan to collaborate with Nos\'e,
who had invited me to visit Keio, I prepared a list of about a dozen
projects on which we might work together, and took it to his office for
discussion.  Still nothing.  In retrospect this turned out well, at
least for me, in that it freed up my time to write another book,
``Computational Statistical Mechanics'', summarizing what I had learned
in my Applied Science teaching at Livermore while enjoying the peaceful
work atmosphere of Japan.  Carol and I spent many a night at the computer
laboratory near Hiyoshi station, working on the manuscript.  Our Hershey
House apartment, overlooking a busy baseball field
was within walking distance so that our working hours weren't limited by
the train schedules. 

At Hershey House we got weekly progress phonecalls from Tony DeGroot, back
in Livermore, who had built a 64-processor parallel computer capable of
million-atom molecular dynamics.  Working with Tony and Jeff Kallman in
Livermore, with the support of Irv Stowers and Fred Wooten,  as well as
collaborating with Taisuke Boku, Toshio Kawai, and Sigeo Ihara in  Japan,
we worked long-distance on color movies of silicon crystal deformation.
A still picture from such a movie is shown here in {\bf Figure 8}.
This collaboration, with nine coauthors, involved the work of more
individuals than did any other of my research efforts\cite{b35}.

\begin{figure}
\includegraphics[height=12cm,width=14cm,angle=-0]{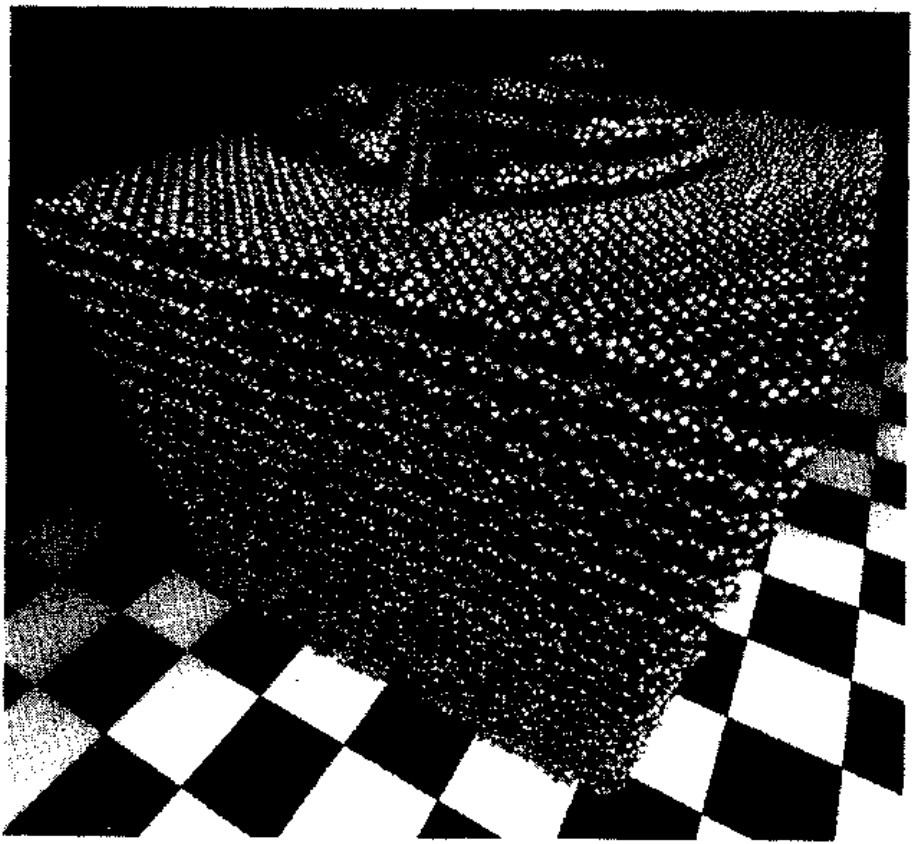}
\caption{
A plastic indentation pit in a 373,248-atom model of silicon.  The indentor
(not shown) is tetrahedral in shape and moves at one-fifth the sound
velocity.  Analogous two-dimensional indentation simulations are
described in Reference 35.
}
\end{figure}

After the Berlin Wall came down in 1989 the Livermore Laboratory was left
without a mission, and became a vestige of its former self, doing ``work
for others'' and striving to appear practical.  In this climate Teller's
Department of Applied Science lost its appeal to the Laboratory, and 
gradually decayed.  Though I continued to work at the Laboratory for
another half dozen years, until a lucrative early-retirement package came
along, the research excitement at Livermore had definitely disappeared.
The working population at the laboratory is now (at the end of 2008) only
half its former size.  My research from about 1994 through to the present
 has been carried out with my wife Carol and a
large international group of collaborators from outside the laboratory.

\section{Lessons Learned from Looking Backward}

My career at Livermore, with its many pleasant overseas interludes,
left me with some powerful lessons, worth outlining here. They have
to do with the value of research and its fruits and how it is best
nurtured.
{\em Publication} of {\em reproducible results} is the {\em sine qua non}
of science.  A research project, no matter how brilliant, is quite useless
unless others can share its results.  Coworkers, administrators, and
editors are the filters through which work must pass before it enters
the literature.  Their suggestions are often good ones.

Some publications are not so good.  George Stell suggested the journal name
``Setbacks in Physics''.  Setbacks in Physics was to be  devoted
exclusively to faulty and incorrect solutions of significant problems
already solved correctly in the literature.  In my own research career
I came across two sets of excellent candidate articles for that journal.
Both candidates were in research areas where I had published extensively.
A series of papers by Tuckerman and coworkers\cite{b36}, pointedly
ignored the simple relationship between Nos\'e-Hoover mechanics and the
Second Law of Thermodynamics due to multifractal phase-space structures
which I had described repeatedly in the literature.  A second, remarkably
long, paper by Zhou\cite{b37} was a specially effective setback,
in that it spawned many successor setback papers, all pushing the original
claim that the (wholly-correct) microscopic form of the Virial Theorem
for the pressure was incorrect, and that
the kinetic part of the pressure tensor should not be included.

A second category of ``bad'' paper, one with forged data, is more rare.
The only example I came across myself was one Watanabe's free-energy
calculation.  His results were literally ``too good to be true''\cite{b38}.
A phonecall to his thesis advisor revealed that the computer program
he used to generate free energies had wild fluctuations.  The program
was simply stopped when the free energy reached the ``correct'' value.

In the past good papers were often squelched by the review process.  I
remember Ed Jaynes had George Uhlenbeck's rejection of Ed's seminal
paper on maximum-entropy theory framed in his Washington University
office.  Such defects of the reviewing process are not so important now.
I have never failed to publish a rejected paper elsewhere.  Because
there are now so many outlets for publication, including the LANL
arXiv and one's own website, arbitrary rejection is no longer a serious
problem.  Today an author can certainly publish if he wishes to do so.

A half-century of simulation work has left me with some lasting lessons.
Reproducibility is paramount, and Clarity is required.  Scepticism and Openness
are desirable, as is also a sense of Perspective.  Visits and discussions with
others most often lead to useful ideas.  Publication is in the end absolutely
necessary despite the occasional frustrations of peer review and the cronyism
that discourages novelty.

For me the Livermore laboratory of the 1960s provided a nearly-ideal
environment for learning about simulation -- stimulating people, freedom
to choose one's own way, plenty of secretarial support and
computing equipment, the possibility of travel and publication.  After
a few years of microscopic simulations, first equilibrium and then
nonequilibrium, as described here, it was natural for me to explore
{\em continuum} methods to get beyond
the limitations on system size and timescale posed by atomistic vibration
lengths and times.  The smooth-particle method, developed by Lucy and
Monaghan in 1977 provides a method much like molecular dynamics for solving
the continuum equations.  But the particles are not necessarily small.  They
can be astrophysical in size.  Smooth particles held my interest for many
years, resulting in my most recent book, which is devoted to that technique.

A Japanese colleague, Shida-san Koichiro, a Lecturer at Musashi Institute
of Technology in Tokyo, had been trained at Keio by Taisuke Boku and
Toshio Kawai.  We all met when Carol and I visited Keio in
1989-1990.  Koichiro was able to work with us at Livermore on Maxwell's
thermal-creep problem.  He has very kindly translated all four of my books into
Japanese.  I am very grateful to him for this and suggest that the reader
interested in more details of my work look for those books, either in 
English or in Japanese: [1] Molecular
Dynamics; [2] Computational Statistical Mechanics; [3] Time Reversibility,
Computer Simulation, and Chaos; [4] Smooth Particle Applied Mechanics, the
State of the Art.  My website  http://williamhoover.info also contains a
wealth of technical information, including electronic forms of books [1],
[2], and [4].

\section{Acknowledgments}

Masaharu Isobe kindly suggested this work and my wife Carol helped prepare
the manuscript.


\begin{thebibliography}{99}


\bibitem{b1}  B. J. Alder and T. E. Wainwright, ``Molecules in Motion'',
              Scientific American {\bf 201}(4), 113-130 (1959).

\bibitem{b2}  F. F. Abraham, ``Two-Dimensional Melting, Solid-State Stability,
              and the Kosterlitz-Thouless-Feynman Criterion'', Physical
              Review B {\bf 23}, 6145-6148 (1981).

\bibitem{b3}  R. W. Zwanzig, ``Virial Coefficients of Parallel Square and
              Parallel Cube Gases'', Journal of Chemical Physics {\bf 24},
              855-856 (1956).

\bibitem{b4}  W. G. Hoover and A. G. De Rocco, ``Sixth Virial Coefficients 
              for Gases of Parallel Hard Lines, Hard Squares, and Hard
              Cubes'', Journal of Chemical Physics, {\bf 34}, 1059-1060 
              (1961).

\bibitem{b5}  W. G. Hoover and J. C. Poirier, ``Determination of Virial
              Coefficients from the Potential of Mean Force'', Journal
              of Chemical Physics {\bf 37}, 1041-1042 (1962). 

\bibitem{b6}  W. W. Wood and J. D. Jacobsen, ``Preliminary Results from
              a Recalculation of the Monte Carlo Equation of State of
              Hard Spheres'', Journal of Chemical Physics {\bf 27},
              1207-1208 (1957).

\bibitem{b7}  B. J. Alder and T. E. Wainwright, ``Phase Transition for a
              Hard Sphere System'', Journal of Chemical Physics {\bf 27},
              1208-1209(1957).

\bibitem{b8}  B. J. Alder, W. G. Hoover, and T. E. Wainwright, 
              ``Cooperative Motion of Hard Disks Leading to Melting'',
              Physical Review Letters {\bf 11}, 241-243 (1963).

\bibitem{b9}  W. G. Hoover and F. H. Ree, ``Melting Transition and
              Communal Entropy for Hard Spheres'', Journal of Chemical
              Physics {\bf 49}, 3609-3617 (1968).

\bibitem{b10} J. B. Gibson, A. N. Goland, M. Milgram, and G. H. Vineyard,
              ``Dynamics of Radiation Damage'', Physical Review {\bf 120},
              1229-1253 (1960).

\bibitem{b11} B. L. Holian, W. G. Hoover, B. Moran, and G. K. Straub,
              ``Shockwave Structure {\em via} Nonequilibrium Molecular 
              Dynamics'', Physical Review A {\bf 22}, 2798-2808 (1980).

\bibitem{b12} O. Kum, Wm. G. Hoover, and C. G. Hoover, ``Temperature
              Maxima in Stable Two-Dimensional Shock Waves'', Physical
              Review E {\bf 56}, 462-465 (1997). 

\bibitem{b13} D. R. Squire, A. C. Holt, and W. G. Hoover, ``Isothermal
              Elastic Constants for Argon.  Theory and Monte Carlo
              Calculations'', Physica {\bf 42}, 388-397 (1969).

\bibitem{b14} W. G. Hoover, A. C. Holt, and D. R. Squire,``Adiabatic
              Elastic Constants for Argon.  Theory and Monte Carlo
              Calculations'', Physica {\bf 44}, 437-443 (1969).

\bibitem{b15} J. A. Barker and D. Henderson, ``What is Liquid?  Understanding
              the States of Matter'', Reviews of Modern Physics {\bf 48}, 
              587-671 (1976).

\bibitem{b16} R. W. Zwanzig, ``Time Correlation Functions and Transport
              Coefficients in Statistical Mechanics'', Annual Review of
              Physical Chemistry {\bf 16}, 67-102 (1965).

\bibitem{b17} W. G. Hoover, B. Moran, R. M. More, and A. J. C. Ladd.
              ``Heat Conduction in a Rotating Disk {\em via} Nonequilibrium
              Molecular Dynamics'', Physical Review A {\bf 24}, 2109-2114
              (1981)

\bibitem{b18} W. T. Ashurst and W. G. Hoover, ``Microscopic Fracture Studies
              in the Two-Dimensional Triangular Lattice'',
              Physical Review B {\bf 14}, 1465-1473 (1976).

\bibitem{b19} W. G. Hoover and W. T. Ashurst, ``Nonequilibrium Molecular
              Dynamics'', Advances in Theoretical Chemistry {\bf 1},
              1-51 (1975).

\bibitem{b20} D. Levesque, L. Verlet, and J. K\"urkijarvi, ``Computer
              Experiments on Classical Fluids. IV. Transport Properties
              and Time-Correlation Functions of the Lennard-Jones Liquid
              Near its Triple Point'', Physical Review A {\bf 7},
              1690-1700 (1973).

\bibitem{b21} W. G. Hoover, ``Adiabatic Hamiltonian Deformation, Linear
              Response Theory, and Nonequilibrium Molecular Dynamics'',
              Lecture Notes in Physics {\bf 132}, 373-380 (1980).

\bibitem{b22} W. G. Hoover, ``Atomistic Nonequilibrium Computer 
              Simulations'', Physica {\bf 118A}, 111-122 (1983).

\bibitem{b23} W. G. Hoover, ``Nonequilibrium Molecular Dynamics at
              Livermore and Los Alamos'', in {\em Microscopic Simulations
              of Complex Hydrodynamic Phenomena}, M. Mareschal and B. L.
              Holian, editors (Plenum Press, New York, 1992).

\bibitem{b24} G. P. F. Ciccotti and W. G. Hoover, Editors, ``Molecular
              Dynamics Simulation of Statistical-Mechanical Systems'',
              Proceedings of the International Enrico Fermi School of
              Physics, Course {\bf 97} (1986).

\bibitem{b25} Wm. G. Hoover, C. G. Hoover, and J. Petravic, ``Simulation
              of Two- and Three-Dimensional Dense-Fluid Shear Viscosities
              {\em via} Nonequilibrium Molecular Dynamics. Comparison of
              Time-and-Space-Averaged Stresses from Homogeneous Doll's
              and Sllod Shear Algorithms with those from Boundary-Driven
              Shear", Physical Review E {\bf 78}, 046701 (2008).

\bibitem{b26} W. G. Hoover, B. Moran, and J. M. Haile, ``Homogeneous
              Periodic Heat Flow {\em via} Nonequilibrium Molecular
              Dynamics'', Journal of Statistical Physics {\bf 37},
              109-121 (1984).

\bibitem{b27} W. G. Hoover, N. E. Hoover, and K. Hanson, ``Exact Hard-Disk
              Free Volumes'', Journal of Chemical Physics {\bf 70},
              1837-1844 (1979).

\bibitem{b28} W. G. Hoover, W. T. Ashurst, and R. Grover, ``Exact
              Dynamical Basis for a Fluctuating Cell Model'',
              Journal of Chemical Physics {\bf 57}, 1259-1262 (1972).

\bibitem{b29} Wm. G. Hoover, C. G. Hoover, and H. A. Posch, ``50 Joint
              Explorations, 1985-2007'', Schr\"odinger Institute (Wien)
              Preprint Archive, $\# 1898$ (2007).

\bibitem{b30} S. Nos\'e, ``A Unified Formulation of the Constant
              Temperature Molecular Dynamics Methods'', Journal of
              Chemical Physics {\bf 81}, 511-519 (1984).

\bibitem{b31} S. Nos\'e, ``A Molecular Dynamics Method for Simulations
              in the Canonical Ensemble'', Molecular Physics
              {\bf 100}, 191-198 (2002).

\bibitem{b32} W. G. Hoover, ``Canonical Dynamics: Equilibrium Phase-Space
              Distributions'', Physical Review A {\bf 31}, 1695-1697 (1985).

\bibitem{b33} H. A. Posch, W. G. Hoover, and F. J. Vesely, ``Canonical
              Dynamics of the Nos\'e Oscillator: Stability, Order, and
              Chaos'', Physical Review A {\bf 33}, 4253-4265 (1986).

\bibitem{b34} B. Moran, W. G. Hoover, and S. Bestiale, ``Diffusion in a
              Periodic Lorentz Gas'', Journal of Statistical Physics
              {\bf 48}, 709-726 (1987).

\bibitem{b35} W. G. Hoover, A. J. De Groot, C. G. Hoover, I. F. Stowers,
              T. Kawai, B. L. Holian, T. Boku, S. Ihara, and J. Belak,
              ``Large-Scale Elastic-Plastic Indentation Simulations
              {\em via} Nonequilibrium Molecular Dynamics'', Physical
              Review A {\bf 42}, 5844-5853 (1990).

\bibitem{b36} Wm. G. Hoover, D. J. Evans, H. A. Posch, B. L. Holian, and
              G. P. Morriss, `Comment on ``Toward a Statistical
              Thermodynamics of Steady States''', Physical Review
              Letters {\bf 80}, 4103-4103 (1998).

\bibitem{b37} M. Zhou, ``A New Look at the Atomic Level Virial Stress --
              On Continuum-Molecular System Equivalence'', Proceedings of
              the Royal Society of London A, {\bf 459}, 2347-2392, (2003).

\bibitem{b38} B. L. Holian, H. A. Posch, and W. G. Hoover, ``Free Energy
              {\em via} Thermostated Dynamic Potential-Energy Changes'',
              Physical Review E {\bf 47}, 3852-3861 (1993).

\end{thebibliography}
\end{document}